\documentclass[secnumarabic,superscriptaddress,amssymb,nobibnotes,aps,prc]{revtex4}

\newcommand{\Pomeron}{I\!\!P}

\usepackage{epsfig}
\usepackage{amsmath}
\usepackage{hyperref}

\begin{document} 

\title{Dynamical model of antishadowing of the nuclear gluon distribution}

\author{L. Frankfurt}
\affiliation{Nuclear Physics Department, School of Physics and Astronomy, Tel Aviv
University, 69978 Tel Aviv, Israel}
\affiliation{Department of Physics, the Pennsylvania State University, State
  College, PA 16802, USA}
\author{V. Guzey}
\affiliation{Petersburg Nuclear Physics Institute (PNPI), National Research Center ``Kurchatov Institute'', Gatchina, 188300, Russia}
\author{M. Strikman}
\affiliation{Department of Physics, the Pennsylvania State University, State
  College, PA 16802, USA}

\date{\today}

\begin{abstract}

We explore the theoretical observation that within the leading twist approximation, the nuclear effects of shadowing and 
antishadowing in non-perturbative nuclear parton distribution functions (nPDFs) at the input QCD evolution scale involve 
diffraction on nucleons of a nuclear target and
originate from merging of two parton ladders belonging to two different nucleons, which are close in the rapidity space.
It allows us to propose that for a given momentum fraction $x_{\Pomeron}$ carried by the diffractive exchange, 
nuclear shadowing and antishadowing should compensate
each other in the momentum sum rule for nPDFs locally on the interval 
$\ln (x/x_{\Pomeron}) \le 1$. We realize this by constructing an explicit model of nuclear 
gluon antishadowing, which has a wide support in $x$, $10^{-4} < x < 0.2$, peaks at $x=0.05-0.1$ at 
the level of $\approx 15$\% for $^{208}$Pb at $Q_0^2=4$  GeV$^2$ and rather insignificantly depends on details of the model. 
We also studied the impact parameter $b$ dependence of antishadowing and found it to be slow.

\end{abstract}

\maketitle

\section{Introduction}
\label{sec:intro}

Three decades of experiments on hard processes with nuclei have established that 
nuclear structure functions are modified compared to their nucleon counterparts, 
for reviews, see~\cite{Frankfurt:1988nt,Arneodo:1992wf,Piller:1999wx}.
Notably, from deep inelastic scattering (DIS) off nuclear targets, there 
emerges the following pattern of nuclear modifications of the ratio of the nucleus to deuteron 
structure functions $R(x,Q^2)=F_2^A(x,Q^2)/F_2^D(x,Q^2)$ at $Q^2$ of the order of a few GeV$^2$ 
($x$ is the Bjorken variable): $R(x,Q^2) < 1$ for  $x \leq 0.05$ (nuclear shadowing), 
$R(x,Q^2) > 1$ for $0.05 < x < 0.2$ (antishadowing), $R(x,Q^2) < 1$ for $0.2 < x  < 0.8$ (the EMC effect), 
and  $R(x,Q^2)> 1$ for $x > 0.8$, which  mostly originates from short-range nucleon correlations in nuclei.

By virtue of the QCD collinear factorization theorem~\cite{Brock:1993sz}, 
nuclear modifications of $R(x,Q^2)$ and of other nuclear observables can be translated  at sufficiently 
large $Q^2$ into modifications of nuclear parton distribution functions (nPDFs)~\cite{deFlorian:2003qf,Hirai:2007sx,Eskola:2009uj,deFlorian:2011fp,Kovarik:2015cma,Khanpour:2016pph,Eskola:2016oht}
characterized by the factor of  $R_j(x,Q^2)=f_{j/A}(x,Q^2)/[Af_{j/N}(x,Q^2)]$, where $f_{j/A}(x,Q^2)$ is the parton (quark or gluon) 
distribution of flavor $j$ in a nucleus, $x$ is the light-cone fraction of the nucleus momentum
carried by parton $j$, and $f_{j/N}(x,Q^2)$ is the parton distribution of a free nucleon.
Note that in practice 
the fits of nPDFs employ the data starting at $Q^2 \sim 1$ GeV$^2$, where higher twist effects at small $x$ could be large and affect the results of the fits.

In this paper, we investigate properties of non-perturbative nPDFs at the input value of the factorization scale of 
$Q_0^2=4$ GeV$^2$ and small $x \leq 0.1$. The magnitude and the shape of $R_j(x,Q_0^2)$ depend on the flavor $j$;
in this work, we follow the trend of $R_j(x,Q_0^2)$ outlined in~\cite{Frankfurt:1988nt,Frankfurt:1990xz}. 
For valence quarks, $R_{q_{\rm val}}(x,Q_0^2)$   closely 
follows $R(x,Q_0^2)$. In the sea quark (antiquark) channel, the only nuclear modification 
is the suppression due to nuclear shadowing ($R_{\bar{q}}(x,Q_0^2) < 1$)  for $x < 0.1$;
for $x > 0.1$, we take $R_{\bar{q}}(x,Q_0^2) = 1$ based on the analysis of nuclear Drell--Yan data indicating
absence of nuclear modifications of sea quark nPDFs in this $x$ region.
In the gluon channel, large nuclear shadowing for $x < 0.05$ is followed by 
sizable antishadowing extending up to $x=0.2$, whose magnitude is constrained by the nPDF momentum sum rule.
For $x > 0.2$, due to lack of constrains,  we assume 
for simplicity
that  $R_g(x,Q_0^2) = 1$
(note that an EMC-like effect may be present for gluons at large $x$).
Indeed, the experimental constraints on $R_g(x,Q^2)$ for $x\ge 0.2$ are very weak so far because the processes of 
dijet and gauge $W$ and $Z$ boson production in $pA$ scattering studied at the CERN Large Hadron Collider (LHC)  
are not sensitive to gluons at such values of $x$, see Figs.~2 and 3 of Ref.~\cite{Eskola:2016oht}.
The BNL Relativistic Heavy Ion Collider (RHIC) data on $\pi^{0}$ production in dAu scattering~\cite{Adler:2006wg,Abelev:2009hx} extends up to $x=0.3$; the nuclear modification 
factor of $R_{dAu}$ in the large-$x$
(large $p_T$) region is consistent with unity within large experimental uncertainties, which indicates that nuclear
modifications of nPDFs are weak in this region.
Note that the ALICE~\cite{Abbas:2013oua,Abelev:2012ba} and CMS~\cite{Khachatryan:2016qhq} data on coherent $J/\psi$ photoproduction in Pb-Pb ultraperipheral collisions at the LHC at $\sqrt{s}=2.76$ TeV 
gave the first direct and weakly model-dependent evidence of large nuclear gluon shadowing down to $x \approx 10^{-3}$~\cite{Guzey:2013xba,Guzey:2013qza}, which agrees very well with the predictions of Refs.~\cite{Eskola:2009uj,Frankfurt:2011cs}.
Note, however, that the predictions of~\cite{Eskola:2009uj} have very large uncertainties, see the discussion in~\cite{Guzey:2013xba}.

The focus of the present work is the antishadowing phenomenon for the gluon nPDF.
We explore the theoretical observation that within the leading twist approximation~\cite{Frankfurt:2011cs}, 
the effects of nuclear shadowing and antishadowing in nPDFs involve diffraction on nucleons of a nuclear target and
originate from merging of two parton ladders belonging to two different nucleons, which are close in the rapidity space. 
We propose a dynamical mechanism for this effect and build an explicit model of the gluon antishadowing by requiring that
the momentum sum rule for nPDFs is satisfied locally in the rapidity space. 
The resulting gluon antishadowing has a wide support in $x$, $10^{-4} < x < 0.2$ and peaks around 
$x=0.05-0.1$ at the level of $\approx 15$\% for $^{208}$Pb, which 
is somewhat smaller than that predicted in the phenomenological approach of~\cite{Frankfurt:2011cs}.

\section{Nuclear shadowing and antishadowing and the momentum sum rule}
\label{sec:shadow}

\subsection{Leading twist nuclear shadowing}
\label{subsec:lt}

In the target rest frame and at moderate energies, nuclear shadowing arises as a consequence of 
multiple interactions of an incoming hadron with nucleons of the target nucleus leading to destructive interference among the 
scattering amplitudes corresponding to the interaction with one, two and more 
nucleons of the target~\cite{Glauber:1955qq}.
At high energies, the physical picture of nuclear shadowing 
changes~\footnote{In the high-energy limit, 
the contribution of planar diagrams corresponding to the shadowing
correction in the Glauber approach cancels exactly~\cite{Mandelstam:1963cw,Gribov:1968fc}.
Thus, nuclear shadowing is given by non-planar diagrams conserving energy--momentum.
Using duality, the sum of non-planar diagrams can be reorganized into a series corresponding 
to the interaction of a projectile with  a given number of nucleons, which superficially has the form 
close to that of the  Glauber multiple scattering series. The major difference is the necessity to include
diffractive intermediate states, i.e., to include color fluctuations. For a review, 
see Ref.~\cite{Frankfurt:2013ria}.}
because
the characteristic longitudinal distance increases with an increase of the projectile momentum and 
becomes comparable to the size of the nuclear target so that the projectile 
(hadron, real and virtual photon, neutrino, vector boson, etc.) interacts with a nuclear target 
by means of its long-lived quark-gluon configurations.  
In the aligned jet model (AJM) for deep inelastic processes off nuclear targets in the target 
rest frame suggested initially within the parton model~\cite{Bjorken:1972uk,Bjorken:1973gc} 
and later generalized to account for QCD phenomena in~\cite{Frankfurt:1988nt}, 
the  interaction of the projectile with a nuclear target can 
be organized in the form of multiple interactions of projectile fluctuations with a given number of target nucleons.
The interaction with $N=2$ nucleons of the target (e.g., the shadowing correction to the 
pion--deuteron total scattering  cross section) can be unambiguously and model-independently 
expressed in terms of the elementary projectile--nucleon diffraction cross section~\cite{Gribov:1968jf}.

\begin{figure}[t]
\begin{center}
\includegraphics[scale=0.8]{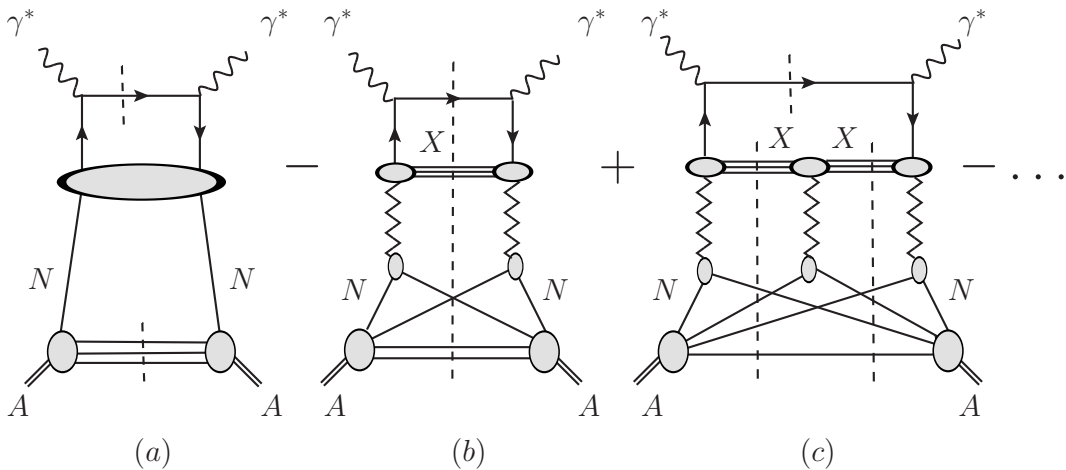}
\caption{The multiple scattering series for the nuclear parton distribution $f_{j/A}(x,Q_0^2)$.}
\label{fig:GG_series}
\end{center}
\end{figure}

\begin{figure}[t]
\begin{center}
\includegraphics[scale=0.45]{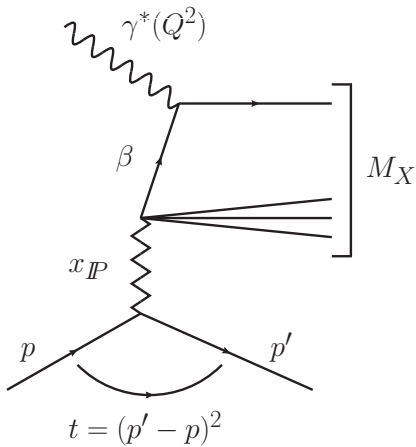}
\caption{
Hard diffractive DIS at the leading order.
}
\label{fig:diffraction_anti}
\end{center}
\end{figure}

In the case of DIS off nuclear targets and nPDFs $f_{j/A}(x,Q_0^2)$, the resulting series of multiple interactions with target
nucleons is shown in Fig.~\ref{fig:GG_series}. These interactions involve diffractive processes, which being a shadow of inelastic ones, are leading twist processes.
As a consequence  of the factorization theorem for diffractive processes~\cite{Collins:1997sr}, 
the shadowing correction to $f_{j/A}(x,Q_0^2)$ originating from the interaction 
with two nucleons of the target can be expressed in terms of the proton diffractive parton
distribution $f_{j/N}^{D}$~\cite{Frankfurt:1998ym}.
At the same time,
the contribution of the interaction with three and more ($N \geq 3$) nucleons of the target 
cannot be model-independently expressed in terms of $f_{j/N}^{D}$. 
However, since it
is dominated by soft (large-size) configurations/components
of the virtual photon, the strength of the interaction with $N \geq 3$ nucleons 
can be parameterized by the effective soft cross section $\sigma_{\rm soft}^j$~\cite{Frankfurt:2011cs}.
This reflects the observation that for the interaction of hadron-like configurations  
(as well as hadrons), fluctuations of the interaction strength lead to very small corrections 
to the total cross section.    Consequently, one can apply
the quasi-eikonal approximation,
which includes diffractive intermediate states, 
 to sum the multiple scattering series for $f_{j/A}(x,Q_0^2)$
in Fig.~\ref{fig:GG_series} and obtain for the shadowing correction 
$\delta xf_{j/A}(x,Q_0^2) \equiv xf_{j/A}(x,Q_0^2)-xf_{j/N}^{\rm IA}(x,Q_0^2)$, where $xf_{j/A}^{\rm IA}(x)$ is 
the nuclear PDF in the impulse approximation:
\begin{eqnarray}
\delta xf_{j/A}(x,Q_0^2) &=&
-8 \pi A (A-1)\, \Re e \frac{(1-i \eta)^2}{1+\eta^2}
\int^{x_0}_{x} d x_{\Pomeron} \beta f_j^{D(4)}(\beta,Q_0^2,x_{\Pomeron},t_{\rm min})\int d^2 b \int^{\infty}_{-\infty}d z_1 \int^{\infty}_{z_1}d z_2
 \nonumber\\
&\times& \rho_A(\vec{b},z_1) \rho_A(\vec{b},z_2) e^{i (z_1-z_2) x_{\Pomeron} m_N}
 e^{-\frac{A}{2} (1-i\eta) \sigma_{\rm soft}^j(x,Q_0^2) \int_{z_1}^{z_2} dz^{\prime} \rho_A(\vec{b},z^{\prime})} \,.
\label{eq:shad_gen}
\end{eqnarray}
In Eq.~(\ref{eq:shad_gen}), $f_j^{D(4)}$ is the diffractive parton distribution of the nucleon
quantifying the parton content of the proton diffractive structure 
function~\cite{Aktas:2006hy,Chekanov:2009qja}, which depends on 
the following variables (Fig.~\ref{fig:diffraction_anti}): $\beta=x/x_{\Pomeron}$; 
$x_{\Pomeron}=(M_X^2+Q_0^2)/(W^2+Q_0^2)$ is the longitudinal momentum fraction loss
by the proton (the light-cone fraction of the diffractive exchange represented by the zigzag line), 
where $M_X$ is the mass of the diffractively-produced final state
and $W^2=(q+p)^2$ is the invariant photon--proton energy squared;
$t=(p^{\prime}-p)^2$ is the momentum transfer squared; $Q_0$ is the initial scale at 
which Eq.~(\ref{eq:shad_gen}) is defined. 
The integration over $x_{\Pomeron}$ corresponds to the sum over diffractively-produced intermediate
states denoted by $X$ in Fig.~\ref{fig:GG_series} and runs from
the minimal kinematically
allowed value of $x$ to $x_0=0.1$, which  
follows directly from the experimentally accessible range of 
$x_{\Pomeron} < 0.1$~\cite{Aktas:2006hy,Chekanov:2009qja}. 
Note that since the large-$x_{\Pomeron}$ contribution is suppressed by the 
$\exp[i (z_1-z_2) x_{\Pomeron} m_N]$ factor,
the exact value of $x_0$ is numerically insignificant.

Further, in Eq.~(\ref{eq:shad_gen}), $\rho_A$ is the nuclear density
parameterized in the standard two-parameter Fermi (Woods--Saxon) form~\cite{DeJager:1987qc}, 
which 
depends on the transverse ($\vec{b}$) and longitudinal ($z$) coordinates of the involved
nucleons; $m_N$ is the nucleon mass; $\eta$ is the ratio of the real to the imaginary parts
of the elementary diffractive $\gamma^{\ast} N \to X N$ scattering amplitude;
the factor of  $\exp[i (z_1-z_2) x_{\Pomeron} m_N]$ 
takes into account the space--time development of the 
process~\cite{Bauer:1977iq}.
Note that all factors of $\rho_A$ enter Eq.~(\ref{eq:shad_gen}) at the same transverse distance $\vec{b}$, 
i.e., all involved nucleons are located at the same impact parameter, since
the $t$ dependence of the elementary virtual photon--nucleon and $X$--nucleon
scattering amplitudes has been neglected compared to that of the nuclear form factor. 
This also explains why  $f_j^{D(4)}$ is evaluated at the minimal momentum transfer 
$t=t_{\rm min} \approx 0$.

Finally, while the exact magnitude of the effective $\sigma_{\rm soft}^j(x,Q_0^2)$ cross section in Eq.~(\ref{eq:shad_gen})
 is model-dependent, its range can be estimated using phenomenological 
information on the hadronic structure (fluctuations) of virtual photons. For instance, in the gluon channel,
the analysis of Ref.~\cite{Frankfurt:2011cs}
gives that $\sigma_{\rm soft}^j(x,Q_0^2)$ is a weak function of $x$ decreasing from $\sigma_{\rm soft}^g(x,Q_0^2)=40-55$ mb at 
$x=10^{-5}$ to $\sigma_{\rm soft}^g(x,Q_0^2)=30-45$ mb at $x=10^{-3}$ and to $\sigma_{\rm soft}^g(x,Q_0^2)=25-40$ mb at $x=0.01$.
The uncertainty in $\sigma_{\rm soft}^j(x,Q_0^2)$ results in an uncertainty spread of predictions for
nuclear shadowing of $f_{j/A}(x,Q_0^2)/[Af_{j/N}(x,Q_0^2)]$ at small $x$.
At the same time, this uncertainty very weakly affects modeling of antishadowing for the gluon nPDF, see our results in Sect.~\ref{sec:xpdep}.

Since the shadowing correction of Eq.~(\ref{eq:shad_gen}) can be in principle defined in terms
of matrix elements of leading twist operators, it is a leading twist quantity. This property is 
explicit in the low nuclear density limit, when the $ N \geq 3$ terms and the associated
$\sigma_{\rm soft}^j$ can be safely neglected and $\delta xf_{j/A}(x,Q_0^2)$ is expressed in terms of the
leading twist diffractive parton distributions of the proton $f_j^{D(4)}$. Thus, the 
$Q^2$ dependence of the resulting nuclear PDFs $f_{j/A}(x,Q^2)$  is given by the usual linear 
Dokshitzer--Gribov--Lipatov--Altarelli--Parisi (DGLAP) evolution.

\subsection{Dynamical approach to the antishadowing phenomenon}
\label{subsec:dm}

Nuclear shadowing in DIS at not too small $x$ is described by an exchange of two ladders.
This is illustrated by graph $b$ of Fig.~\ref{fig:GG_series} in the target rest frame (each zigzag line represents
a ladder).

\begin{figure}[t]
\begin{center}
\includegraphics[scale=1.0]{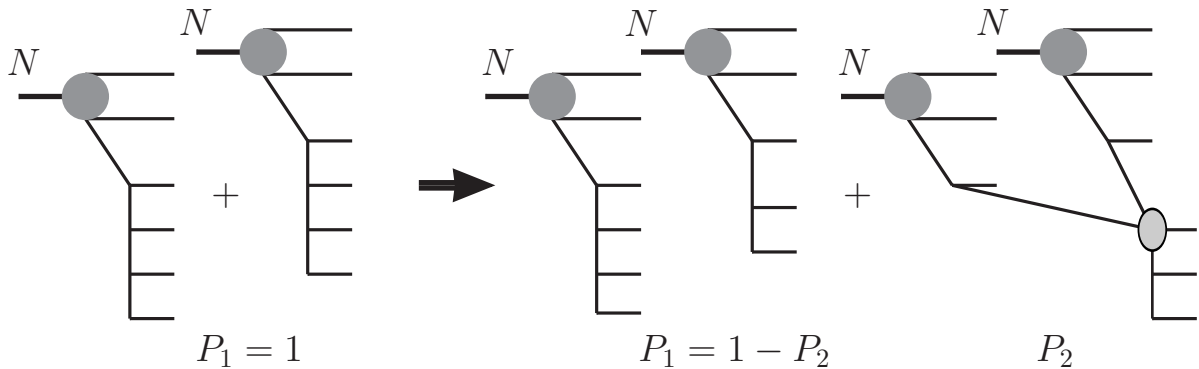}
\caption{Merging of two ladders coupled to two different nucleons in the $2\Pomeron \to \Pomeron$ 
process in the nucleus infinite momentum
frame. This process corresponds both to nuclear shadowing and antishadowing.
}
\label{fig:antishadowing_IMF}
\end{center}
\end{figure}

In the triple Pomeron limit approximation, which is consistent with the HERA data on hard inclusive diffraction in
$ep$ DIS~\cite{Aktas:2006hy,Chekanov:2009qja}, 
this contribution can be considered as a result of 
 emission of two ladders at different impact parameters~\cite{Gribov:1973jg}. Partons of these two ladders may come close 
 together in the impact parameter plane due to diffusion and merge into one ladder.  In the infinite momentum frame (IMF),
  this corresponds to a reduction of the probability for a fast nucleus (deuteron) to be in the configuration,  
  where its small $x$ component is described as a system of two independent ladders originating from two nucleons, 
  and an additional contribution to the wave function, where the system is described by 
  two ladders for the values of the rapidity below the rapidity, where the merger occurred, 
  see Fig.~\ref{fig:antishadowing_IMF}.
  As a result, at given small $x$,
 the probability  
 to have 
 two independent ladders 
is given by the probability of diffraction in a given channel; we denote this probability $P_1$.
The probability that merging occurs above given $x$ is $P_2=1- P_1 $. Obviously,
 for large $x$ this model corresponds to nPDFs being equal to the sum of individual nucleon PDFs, 
while for small $x$, the relative reduction of nPDFs is given by the factor of $P_2$. Note that the process illustrated in Fig.~\ref{fig:antishadowing_IMF}
is analogous but not identical to the familiar triple Pomeron processes
in hadronic collisions. 
(Note that the third ladder may be rather short and not be described by a Pomeron exchange.) 

In the nucleus IMF, the merging shown in the right graph in Fig.~\ref{fig:antishadowing_IMF} 
means that a fraction of the nucleus 
momentum carried by the third ladder is a sum of the momentum fractions taken from the two ladders. 
Therefore, after the merging of the ladders, the fractions of target momentum are larger than within a single ladder. 
Hence, the contribution of the  diagrams presented at Fig.~\ref{fig:antishadowing_IMF} to nPDFs   
is positive  at larger $x$.  For sufficiently small $x$, these diagrams produce a negative contribution to nPDFs, i.e., they
lead to nuclear shadowing.
It is essential to point out that 
since the graphs in Fig.~\ref{fig:antishadowing_IMF} conserve energy--momentum, they represent
a sum of the nuclear shadowing and antishadowing contributions and allow us to formulate a dynamical
approach to the antishadowing phenomenon. 

The next important observation is that the QCD analysis of the HERA diffractive data~\cite{Aktas:2006hy,Chekanov:2009qja} 
indicates that diffraction in DIS is dominated by soft Pomeron-like interactions, which follows from
the observation that  $\alpha_{\Pomeron}(0)$ in DIS is practically the  same as for soft interactions.
Since in soft interactions the correlation length in rapidity $\Delta y  \sim 1$, 
modifications of parton densities related to the merging of the two ladders should be rather local in
 the rapidity and located close to the rapidity position of the vertex describing the 
 $2\Pomeron \to \Pomeron$ [$(n \Pomeron) \to \Pomeron$]
transition.  Therefore,  for a given light-cone momentum $x_{\Pomeron}$ carried by the lower ladder in 
Fig.~\ref{fig:antishadowing_IMF}, the merging of ladders should predominantly correspond to   
$\ln (x/x_{\Pomeron}) \le 1$. This means that for a given $x_{\Pomeron}$, nuclear shadowing and
 antishadowing should compensate
each other in the momentum sum rule for nPDFs on the interval $\ln (x/x_{\Pomeron}) \le 1$.

While the lack of the detailed knowledge of the parton structure of the $2\Pomeron \to \Pomeron$  
vertex does not allow us to built a microscopic theory of antishadowing, the realization of 
the observation that the momentum sum rule is valid locally on the $\ln (x/x_{\Pomeron}) \le 1$ 
interval  enables us to model antishadowing with only  
modest uncertainty in the final results. 

Above we discussed the dynamical model of shadowing and antishadowing originating from an exchange of two ladders belonging
to two different nucleons of the nucleus, which exhausts the answer in the cases of low nuclear density and the deuteron.
In a general case, one needs to take into account the interaction with $N \geq 3$ nucleons of the nucleus, which can be done using the quasi-eikonal approximation with the effective cross section $\sigma^j_{\rm soft}$, see Eq.~(\ref{eq:shad_gen}).
These additional elastic interactions do not involve the ``first'' and the ``last'' nucleons, which couple to the merging ladders
(see Fig.~\ref{fig:GG_series}),
and, hence, do not affect the general picture shown in Fig.~\ref{fig:antishadowing_IMF}.

Note that in this paper, we are concerned with the gluon nPDF at small $x$, i.e., shadowing and 
antishadowing in the vacuum channel. The dynamics of shadowing and antishadowing in the non-vacuum channel 
relevant for valence quark nPDFs and polarized nPDFs involves the Pomeron--Reggeon interference and merging, which is 
characterized by smaller diffractive masses (the $P_2$ probability) and a combinatoric enhancement of the shadowing 
term. This in general results in the $x$ dependence and magnitude of shadowing and antishadowing in the non-vacuum channel which
are different from those in the gluon channel.

Shadowing and antishadowing of antiquark nPDFs and the nuclear structure function $F_{2A}(x,Q^2)$ 
at moderate $Q^2$ was modeled in 
Ref.~\cite{Brodsky:1989qz} using the Glauber theory and the high-energy Regge behavior of the antiquark--nucleon scattering
amplitude $T_{\bar{q}N}$. In this approach, an antishadowing enhancement arises due to the real part of $T_{\bar{q}N}$, 
which is given by the $\alpha_R=1/2$ Reggeon exchanges. 
Note shadowing and antishadowing in the gluon channel were not considered in~\cite{Brodsky:1989qz}.

\subsection{Constraining antishadowing using the momentum sum rule including Coulomb effects}
\label{subsec:anti}

We discussed in the Introduction that in the gluon channel,  the small $x$ shadowing effect 
should be supplemented by the effect of antishadowing, which peaks at intermediate $x$ ($0.1 < x < 0.2$). 
To constrain the gluon antishadowing, one can use the momentum sum rule for nuclear PDFs:
\begin{equation}
\frac{1}{A}\sum_j \int_0^{A} dx x f_{j/A}(x,Q^2)=1-\eta_{\gamma}(A) \,, 
\label{eq:mom_sr}
\end{equation}
where the sum runs over all flavors $j$; $x=A Q^2/2 (q \cdot P_A)$ is the rescaled Bjorken 
$x$ ($ 0 < x < A$), where 
$q$ and $P_A$ are the four-momenta of the virtual photon and the nucleus, respectively;
$\eta_{\gamma}(A)$ is the momentum fraction of a fast moving nucleus carried by equivalent 
photons, which is of the order of a fraction of the percent for heavy nuclei~\cite{Frankfurt:2012qs}.

Nuclear modifications of $f_{j/A}(x,Q^2)$ change its shape with respect to the the impulse approximation.
Since the discussed effects are not large, it is necessary to take into
account the momentum carried by  equivalent photons (explicit nucleus 
non-nucleonic degrees of freedom) in the  nucleus wave function. For the impulse approximation (IA), one obtains:
\begin{equation}
xf_{j/A}^{\rm IA}(x,Q^2)=  \left[Z x_p^{\prime} f_{j/p}(x_p^{\prime})+N x_p^{\prime}f_{j/n}(x_p^{\prime})\right] \,,
\label{eq:npdf_IA}
\end{equation}
where $Z$ is the nucleus charge; $N$ is the number of neutrons; $x_{p}^{\prime}=x_p/(1-\eta_{\gamma}(A))$ 
and  $x_p=Q^2/(2 p \cdot q)$. The rescaling $x_p \to x_{p}^{\prime}$~\cite{Frankfurt:2012qs} enables one 
to satisfy the momentum sum rule~(\ref{eq:mom_sr}) for $f_{j/A}^{\rm IA}(x,Q^2)$:
\begin{equation}
\frac{1}{A}\sum_j \int_0^{A} dx x f_{j/A}^{\rm IA}(x,Q^2)=\frac{Z}{A} \sum_j \int_0^{1} dx_p x_p^{\prime} f_{j/p}(x_p^{\prime},Q^2)+\frac{N}{A} \sum_j \int_0^{1} dx_p x_p^{\prime} f_{j/n}(x_p^{\prime},Q^2)=1-\eta_{\gamma}(A) \,.
\label{eq:mom_sr_IA}
\end{equation}
Therefore, the momentum sum rule~(\ref{eq:mom_sr}) can be rewritten in the following form:
\begin{equation}
\sum_j \int_0^{1} dx x \left[f_{j/A}(x,Q^2)-f_{j/A}^{\rm IA}(x,Q^2)\right]=0 \,.
\label{eq:mom_sr_2}
\end{equation}
In Eq.~(\ref{eq:mom_sr_2}) we neglected the contribution of the $x > 1$ region, which is expected to be 
numerically insignificant.

Writing explicitly the sum over parton flavors, using Eq.~(\ref{eq:shad_gen}) for the shadowing
correction in the sea quark and gluon channels and introducing the gluon antishadowing contribution
$\delta x g_A^{\rm anti}(x,Q_0^2)$ for $x< 0.2$, 
Eq.~(\ref{eq:mom_sr_2}) at $Q^2=Q_0^2$ can be rewritten in the following form:
\begin{equation}
2\sum_{q=u,d,s,c} \int_0^{0.1} dx \delta x {\bar q}_A(x,Q_0^2)+\int_0^{0.1} dx \delta x g_A(x,Q_0^2) 
+\int_0^{0.2} dx \delta x g_A^{\rm anti}(x,Q_0^2)=0 \,.
\label{eq:mom_sr_3}
\end{equation}
Note that the contribution 
of valence quarks to Eq.~(\ref{eq:mom_sr_3}), whose medium modifications are constrained by the baryon 
number sum rule, is numerically smaller by approximately an order of magnitude than each of the shown terms of 
this equations and, hence, has been safely neglected.

While the shape of the gluon antishadowing is unknown, it is also a coherent nuclear effect
as follows from the diagrams discussed above.
Coherent nuclear effects rapidly vanish, when the coherence length $l_c \approx 1/(2 m_N x)$ becomes 
comparable to the average distance between two nucleons in a nucleus, $r_{NN} \approx 1.7$ fm. 
This corresponds to $x \approx 0.2$ in the momentum space and we use this value as an upper limit on
the antishadowing support in Eq.~(\ref{eq:mom_sr_3}).

Equation~(\ref{eq:mom_sr_3}) constrains the first moment of the gluon antishadowing contribution 
and, hence, can be used to model its shape. For instance, assuming 
for the illustration
that $\delta x g_A^{\rm anti}(x,Q_0^2)=R_g^{\rm anti}(x)x g_p(x,Q_0^2)$ on the $0.03 < x < 0.2$ interval and
$\delta x g_A^{\rm anti}(x,Q_0^2)=0$, when $x$ is beyond this interval, and parameterizing  
$R_g^{\rm anti}(x)=N^{\rm anti} (x-0.03) (0.2-x)$, one finds for $^{208}$Pb that $N^{\rm anti} \approx 30$,
which corresponds to $\approx 20$\% enhancement of the $g_A(x,Q_0^2)/[A g_N(x,Q_0^2)]$ 
ratio near $x=0.1$~\cite{Frankfurt:2011cs}.

A similar shape and magnitude of gluon antishadowing was first suggested in \cite{Frankfurt:1990xz} based on the QCD aligned jet model of the leading twist nuclear shadowing and the momentum sum rule. 
It is also similar to the central value --- but with much smaller uncertainties --- of gluon antishadowing
obtained in the EPS09 and 
EPPS16 global QCD analyses of nuclear 
PDFs~\cite{Eskola:2009uj,Eskola:2016oht}.
 It is also important to note that the EPS09 predictions for the gluon 
antishadowing are in the good agreement with the LHC data on the shape of the dijet
pseudorapidity distributions measured by the CMS collaboration in proton--lead collisions at 
$\sqrt{s}=5.02$ TeV~\cite{Paukkunen:2014nqa}.
It was recently confirmed by the EPPS16 analysis~\cite{Eskola:2016oht} using the CMS dijet $pA$ data in the fit.
Note that some enhancement of the nuclear gluon distribution on the interval $0.05 < x < 0.15$ is 
suggested by the NMC data on $J/\psi$ production in deep inelastic
muon scattering on Sn and C nuclei~\cite{Amaudruz:1991sr}.

At the same time, other global QCD analyses suggest completely different shapes of the gluon 
antishadowing indicating that the fixed-target data does not constrain it. For instance, in the HKN07 
analysis~\cite{Hirai:2007sx},
the gluon antishadowing starts at $x \approx 0.1$ and rapidly grows as $x$ increases. 
In the DSSZ analysis~\cite{deFlorian:2011fp},
both nuclear shadowing and antishadowing in the gluon channel are small, order of a few percent, effects.

\section{Realization of the dynamical approach to antishadowing}
\label{sec:xpdep}

In the absence of a dynamical model of antishadowing, Eq.~(\ref{eq:mom_sr_3}) is essentially the only
constraint on antishadowing with the ensuing ambiguity mentioned in Sect.~\ref{subsec:anti}.
However, the observation that antishadowing compensates nuclear shadowing in the momentum sum 
rule locally on the $\ln (x/x_{\Pomeron}) \le 1$ interval (see Sect.~\ref{subsec:dm}) allows us to build a more detailed model.

To realize the dynamical approach to antishadowing,
we introduce the gluon antishadowing 
$\delta x g_A^{\rm anti}(x,x_{\Pomeron},Q_0^2)$, which depends both on $x$ and $x_{\Pomeron}$
and is normalized by the following relation:
\begin{equation}
\int_0^{0.1} dx_{\Pomeron}  \delta x g_A^{\rm anti}(x,x_{\Pomeron},Q_0^2)=\delta x g_A^{\rm anti}(x,Q_0^2) \,,
\label{eq:anti_xp1}
\end{equation} 
where $\delta x g_A^{\rm anti}(x,Q_0^2)$ enters Eq.~(\ref{eq:mom_sr_3}).
This allows us to explicitly introduce the integration over $x_{\Pomeron}$ in 
the momentum sum rule~(\ref{eq:mom_sr_3}):
\begin{equation}
\sum_{j} \int_0^{0.1} dx  \int_{x}^{0.1} dx_{\Pomeron} \delta x f_{j/A}(x,x_{\Pomeron},Q_0^2)
+\int_0^{0.2} dx \int_{0}^{0.1} dx_{\Pomeron} \delta x g_A^{\rm anti}(x,x_{\Pomeron},Q_0^2)=0 \,,
\label{eq:anti_xp2}
\end{equation}
where $\sum_{j}$ is the sum over sea quarks and gluons;
$\delta x f_{j/A}(x,x_{\Pomeron},Q_0^2)$ is the shadowing correction as a function of $x$ and 
$x_{\Pomeron}$, which builds up the $x_{\Pomeron}$-integrated shadowing contribution 
in Eq.~(\ref{eq:shad_gen}):
\begin{eqnarray}
\delta xf_{j/A}(x,x_{\Pomeron},Q_0^2) &=&
-8 \pi A (A-1)\, \Re e \frac{(1-i \eta)^2}{1+\eta^2}
 \beta f_j^{D(4)}(\beta,Q_0^2,x_{\Pomeron},t_{\rm min})\int d^2 b \int^{\infty}_{-\infty}d z_1 \int^{\infty}_{z_1}d z_2
 \nonumber\\
&\times& \rho_A(\vec{b},z_1) \rho_A(\vec{b},z_2) e^{i (z_1-z_2) x_{\Pomeron} m_N}
 e^{-\frac{A}{2} (1-i\eta) \sigma_{\rm soft}^j(x,Q_0^2) \int_{z_1}^{z_2} dz^{\prime} \rho_A(\vec{b},z^{\prime})} \,.
\label{eq:anti_xp3}
\end{eqnarray}

Changing the order of integration over $x$ and  $x_{\Pomeron}$ in Eq.~(\ref{eq:anti_xp2}), we then 
require that it is satisfied at each $x_{\Pomeron}$:
\begin{equation}
\sum_{j} \int_0^{x_{\Pomeron}} dx   \delta x f_{j/A}(x,x_{\Pomeron},Q_0^2)
+\int_0^{0.2} dx \delta x g_A^{\rm anti}(x,x_{\Pomeron},Q_0^2)=0 \,.
\label{eq:anti_xp4}
\end{equation}
Equation~(\ref{eq:anti_xp4}) realizes the dynamical approach to antishadowing and leads to 
the constraints on the gluon antishadowing, which are 
more stringent and detailed than those given by Eq.~(\ref{eq:mom_sr_3}).

\begin{figure}[h]
\begin{center}
\includegraphics[scale=0.8]{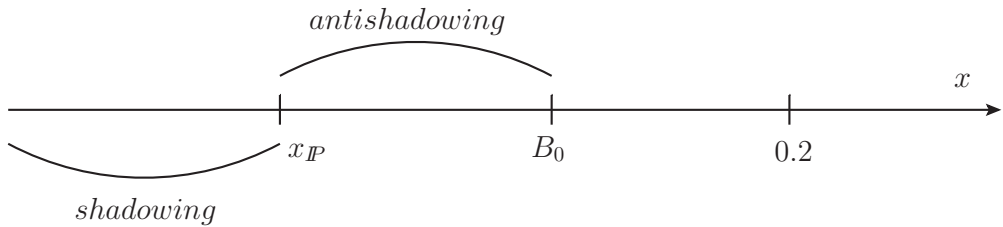}
\caption{Pattern of $x$ and $x_{\Pomeron}$ dependence of the gluon
shadowing and antishadowing.}
\label{fig:anti_sketch}
\end{center}
\end{figure}

To model the gluon antishadowing using Eq.~(\ref{eq:anti_xp4}), we assume that for each 
$x_{\Pomeron}$, $\delta x g_A^{\rm anti}(x,x_{\Pomeron},Q_0^2)$ has support on the interval 
$x_{\Pomeron} \leq x \leq B_0\leq 0.2$ (see Fig.~\ref{fig:anti_sketch}) and is parameterized in the
following simple form:\begin{equation}
\delta x g_A^{\rm anti}(x,x_{\Pomeron},Q_0^2)=\left\{\begin{array}{ll}
N^{\rm anti}(x_{\Pomeron}) (\ln x-\ln x_{\Pomeron}) (\ln B_0-\ln x) xg_N(x,Q_0^2) \,, &  x_{\Pomeron} \leq x \leq B_0\,, \\
0 \,, & x<x_{\Pomeron},\,  x>B_0 \,.
\end{array} \right.
\label{eq:form_b}
\end{equation}
The parameter $B_0$ determines how local in $x/x_{\Pomeron}$ the antishadowing contribution is. 
The $\ln (x/x_{\Pomeron}) \le 1$ condition corresponds to $B_0 \leq 3 x_{\Pomeron}$;
in our analysis, we used $B_0=3 x_{\Pomeron}$ ($B_0 \leq 0.2$) corresponding 
the rapidity merging range of $\Delta y = 1$. We also found that our results very weakly depend on the explicit
value of $B_0$ in the $B_0=3x_{\Pomeron}-5 x_{\Pomeron}$ interval.
The parameter 
$N^{\rm anti}(x_{\Pomeron})$ is determined from Eq.~(\ref{eq:anti_xp4}).

Following our analysis in Ref.~\cite{Frankfurt:2011cs},
for the gluon distribution of the free nucleon, we used the NLO CTEQ5M parametrization~\cite{Lai:1999wy}.
The sensitivity of $g_A(x,Q_0^2)/[Axg_N(x,Q_0^2)]$ to the used underlying free nucleon PDFs was studied 
in~\cite{Frankfurt:2011cs} and it was found that, for instance, the difference between the CTEQ5M and CTEQ66 parametrizations 
affects $g_A(x,Q_0^2)/[Axg_N(x,Q_0^2)]$ only for $x < 10^{-3}$ leading 
to at most a
25\% difference at  $x=10^{-4}$, see Fig.~49 of~\cite{Frankfurt:2011cs}. 
As we already mentioned in Sect.~\ref{subsec:lt}, uncertainties of this magnitude in the gluon nPDF at very small $x$
do not noticeably affect our modeling of the gluon antishadowing as well as the momentum sum rule, see our results in Sect.~\ref{sec:xpdep}.

The sketch of the assumed pattern of the $x$ and $x_{\Pomeron}$ dependence of 
$\delta x g_A^{\rm anti}(x,x_{\Pomeron},Q_0^2)$ is shown in Fig.~\ref{fig:anti_sketch}.
 
\begin{figure}[h]
\begin{center}
\includegraphics[scale=1.7]{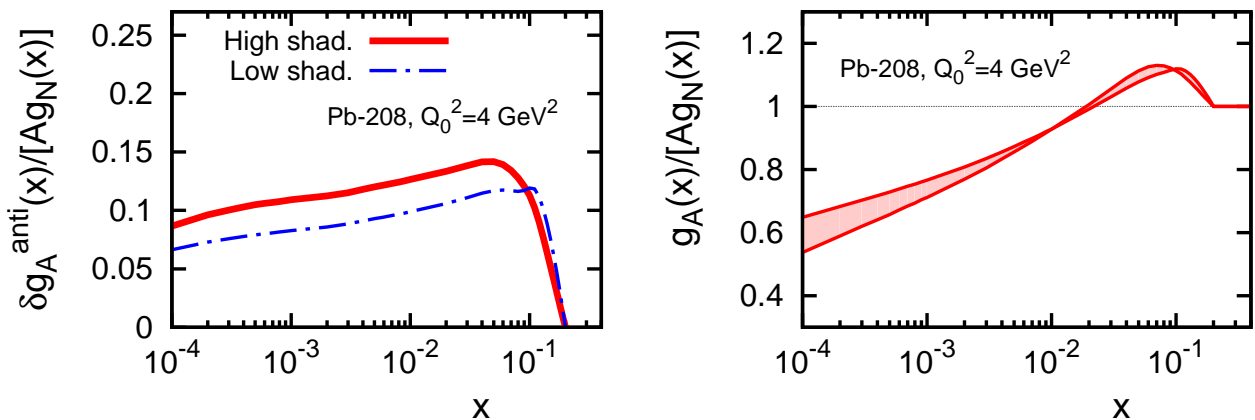}
\caption{$\delta x g_A^{\rm anti}(x,Q_0^2)/[Axg_N(x,Q_0^2)]$ (left) and $x g_A(x,Q_0^2)/[Axg_N(x,Q_0^2)]$ (right) as a function of
$x$ for $^{208}$Pb at $Q_0^2=4$ GeV$^2$. See text for details.
}
\label{fig:anti_sum_old_xdep_2014}
\end{center}
\end{figure}

Figure~\ref{fig:anti_sum_old_xdep_2014} (left) presents our results for 
$\delta x g_A^{\rm anti}(x,Q_0^2)/[Axg_N(x,Q_0^2)]$ as a function of
$x$ for $^{208}$Pb at $Q_0^2=4$ GeV$^2$. 
The solid and dot-dashed curves labeled ``High shad.'' and ``Low shad.'' correspond 
to the scenarios with the higher and lower nuclear gluon shadowing~\cite{Frankfurt:2011cs}, respectively. 
One can see from the figure that in all cases, the antishadowing enhancement does not exceed 15\%
and peaks around $x \approx 0.05 - 0.1$.  Note also that the effect of antishadowing is rather small for $x \leq 10^{-4}$. This is a consequence 
of the fact that for these values of $x$, the shadowing correction---and, hence, the compensating 
antishadowing contribution---receives the dominant contribution from
the intermediate diffractive masses corresponding to $x_{\Pomeron} \geq 10^{-4}$.

Figure~\ref{fig:anti_sum_old_xdep_2014} (right) 
presents our predictions for 
$x g_A(x,Q_0^2)/[Axg_N(x,Q_0^2)]$ as a function of
$x$ for $^{208}$Pb at $Q_0^2=4$ GeV$^2$. The shaded band spans the range of our predictions for the gluon nuclear shadowing~\cite{Frankfurt:2011cs}
and antishadowing. 
Note that in this work we present our results for $x g_A(x,Q_0^2)/[Axg_N(x,Q_0^2)]$ for  $x> 10^{-4}$, 
where the data on diffraction in $ep$ scattering are
available from HERA. Extrapolation of the HERA fits to smaller $x$ allows one to make
estimates for nuclear shadowing for even smaller smaller $x$, see Fig.~ 31 in Ref.~\cite{Frankfurt:2011cs}.

\begin{figure}[th]
\begin{center}
\includegraphics[scale=1.4]{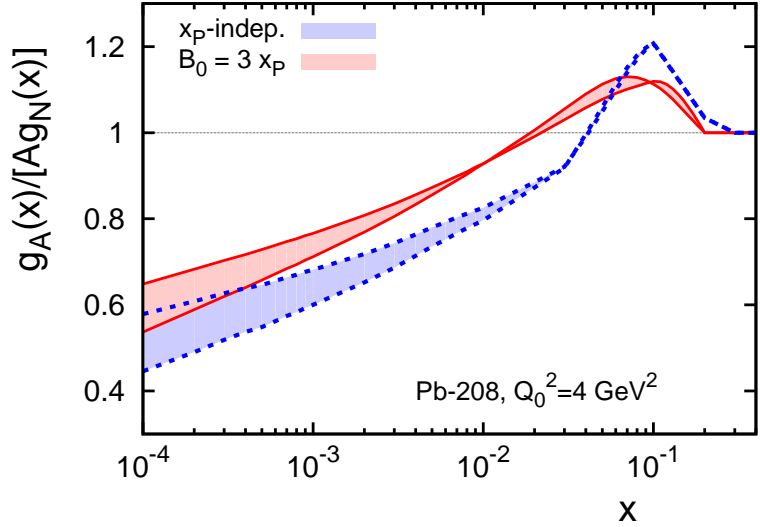}
\caption{Comparison of $x g_A(x,Q_0^2)/[Axg_N(x,Q_0^2)]$, when antishadowing is 
modeled using Eqs.~(\ref{eq:anti_xp4}) and (\ref{eq:form_b}) 
with $B_0=3 x_{\Pomeron}$ (the upper shaded band) 
with the case  when it is modeled using Eq.~(\ref{eq:mom_sr_3}) 
(the lower shaded band labeled ``$x_{\Pomeron}$-indep.''). 
}
\label{fig:Santi_sum_old_xdep_2014_comp}
\end{center}
\end{figure}

In Fig.~\ref{fig:Santi_sum_old_xdep_2014_comp}, we compare our predictions for 
$x g_A(x,Q_0^2)/[Axg_N(x,Q_0^2)]$, when antishadowing is modeled as described in this work using
Eqs.~(\ref{eq:anti_xp4}) and (\ref{eq:form_b}) with $B_0=3 x_{\Pomeron}$ (the upper shaded band) 
with the case  when it is modeled 
using the momentum sum rule of Eq.~(\ref{eq:mom_sr_3}) as was done 
in~\cite{Frankfurt:2011cs} (the lower shaded band labeled ``$x_{\Pomeron}$-indep.''). 
One can see from the figure that for small $x < 10^{-4}$, the predictions of the two approaches 
are very close in agreement with the small gluon antishadowing in this region of $x$ as
shown in Fig.~\ref{fig:anti_sum_old_xdep_2014}. 
As $x$ is increased, the contribution of antishadowing in the $B_0=3 x_{\Pomeron}$ case, which has
a wide support in $x$, increases  $x g_A(x,Q_0^2)/[Axg_N(x,Q_0^2)]$. 
At the same time, by construction, 
the $x_{\Pomeron}$-independent antishadowing has the support only for $x \geq 0.03$ and, hence, does 
not affect $x g_A(x,Q_0^2)/[Axg_N(x,Q_0^2)]$ for $x < 0.03$. The difference between the two 
approaches for $10^{-4} < x < 0.05$ is clearly seen in the figure.
Finally, since in the $B_0=3 x_{\Pomeron}$ case the gluon antishadowing has a wider support in $x$ 
than in the $x_{\Pomeron}$-independent case, at $x \approx 0.05 -0.1$ 
it peaks at the level of $\approx 15$\%, which is somewhat smaller than the $20$\% antishadowing
enhancement in the $x_{\Pomeron}$-independent case.

The procedure described above can also be applied to model the impact parameter $b$ dependence
of antishadowing. Removing the integration over $b$ in Eq.~(\ref{eq:anti_xp3}) and thus introducing
the shadowing correction at each $b$, $\delta xf_{j/A}(x,b,x_{\Pomeron},Q_0^2)$:
\begin{eqnarray}
\delta xf_{j/A}(x,b,x_{\Pomeron},Q_0^2) &=&
-8 \pi A (A-1)\, \Re e \frac{(1-i \eta)^2}{1+\eta^2}
 \beta f_j^{D(4)}(\beta,Q_0^2,x_{\Pomeron},t_{\rm min}) \int^{\infty}_{-\infty}d z_1 \int^{\infty}_{z_1}d z_2
 \nonumber\\
&\times& \rho_A(\vec{b},z_1) \rho_A(\vec{b},z_2) e^{i (z_1-z_2) x_{\Pomeron} m_N}
 e^{-\frac{A}{2} (1-i\eta) \sigma_{\rm soft}^j(x,Q_0^2) \int_{z_1}^{z_2} dz^{\prime} \rho_A(\vec{b},z^{\prime})} \,,
\label{eq:anti_impact1}
\end{eqnarray}
we require that the momentum sum rule differential in $x_{\Pomeron}$ [Eq.~(\ref{eq:anti_xp4})] 
is satisfied also at each $b$:
\begin{equation}
\sum_{j} \int_0^{x_{\Pomeron}} dx   \delta x f_{j/A}(x,b,x_{\Pomeron},Q_0^2)
+\int_0^{0.2} dx \delta x g_A^{\rm anti}(x,b,x_{\Pomeron},Q_0^2)=0 \,.
\label{eq:anti_impact2}
\end{equation}
The impact parameter dependent gluon antishadowing $\delta x g_A^{\rm anti}(x,b,x_{\Pomeron},Q_0^2)$ 
can be modeled analogously to Eq.~(\ref{eq:form_b}) in the following form:
\begin{equation}
\delta x g_A^{\rm anti}(x,b,x_{\Pomeron},Q_0^2)=\left\{\begin{array}{ll}
N^{\rm anti}(b,x_{\Pomeron}) T_A(b)(\ln x-\ln x_{\Pomeron}) (\ln B_0-\ln x) xg_N(x,Q_0^2) \,, &  x_{\Pomeron} \leq x \leq B_0\,, \\
0 \,, & x<x_{\Pomeron},\,  x>B_0 \,,
\end{array} \right.
\label{eq:impact3}
\end{equation}
where $T_A(b)=\int dz \rho_A(\vec{b},z)$ is the nuclear optical density; the coefficients 
$N^{\rm anti}(b,x_{\Pomeron})$ are found from Eq.~(\ref{eq:anti_impact2}).
Finally, the resulting antishadowing as a function of $b$ is found after the integration of 
$\delta x g_A^{\rm anti}(x,b,x_{\Pomeron},Q_0^2)$ over $x_{\Pomeron}$:
\begin{equation}
\delta x g_A^{\rm anti}(x,b,Q_0^2)=\int_0^{0.1} dx_{\Pomeron}  \delta x g_A^{\rm anti}(x,b,x_{\Pomeron},Q_0^2) \,.
\label{eq:anti_impact4}
\end{equation}

\begin{figure}[th]
\begin{center}
\includegraphics[scale=1.4]{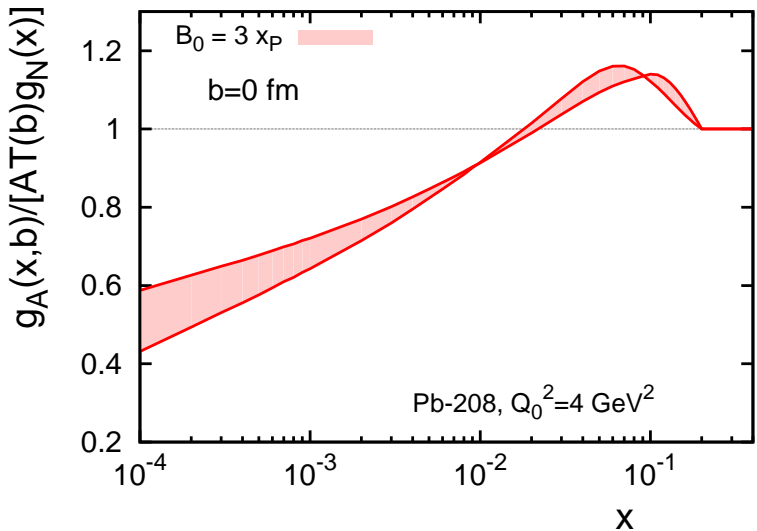}
\caption{Impact parameter dependent gluon nuclear distribution of $^{208}$Pb at $Q_0^2=4$ GeV$^2$.
The $x g_A(x,b,Q_0^2)/[AT_A(b)xg_N(x,Q_0^2)]$ ratio as a function of $x$ at $b=0$ in 
the dynamical approach to antishadowing.
}
\label{fig:Santi_sum_old_xdep_2014_B}
\end{center}
\end{figure}

An example of the resulting impact parameter dependent nuclear gluon distribution is presented in
Fig.~\ref{fig:Santi_sum_old_xdep_2014_B}, where for the $B_0=3 x_{\Pomeron}$ case,
we show $x g_A(x,b,Q_0^2)/[AT_A(b)xg_N(x,Q_0^2)]$
as a function of $x$ for $^{208}$Pb at $Q_0^2=4$ GeV$^2$ and for the central impact parameter 
$b=0$. A comparison with the right panel of Fig.~\ref{fig:anti_sum_old_xdep_2014} shows that while
 the nuclear shadowing effect
noticeably increases as one decreases $b$, this has a much smaller effect on antishadowing, which increases
only by a few percent as one goes from the $b$-integrated case to the $b=0$ case. This is a consequence of 
the fact that in the dynamical approach to antishadowing, it has a wide support in $x$.

\begin{figure}[h]
\begin{center}
\includegraphics[scale=1.2]{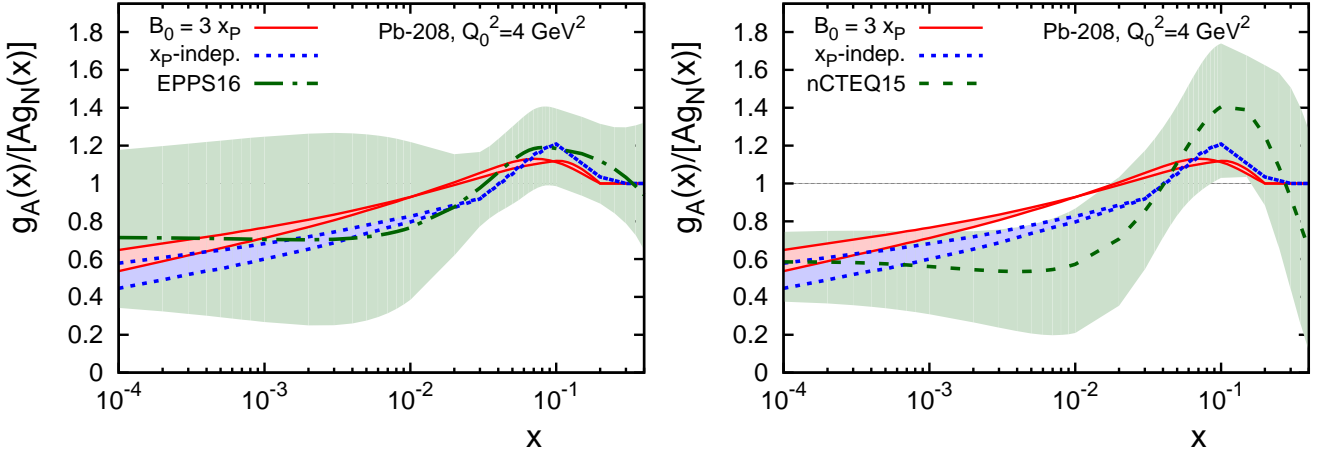}
\caption{Comparison of the prediction of the leading twist nuclear shadowing and the dynamical model of antishadowing for 
$x g_A(x,Q_0^2)/[Axg_N(x,Q_0^2)]$ (same as in Fig.~\ref{fig:Santi_sum_old_xdep_2014_comp}) with results of the EPPS16 (left panel) and nCTEQ15 (right panel) fits. The shaded error bands around the EPPS16 and nCTEQ15 curves give their uncertainties.}
\label{fig:Comp_EPS_2017_v2}
\end{center}
\end{figure}

Our predictions for the leading twist nuclear shadowing and the dynamical model of antishadowing of gluon nPDFs
can be compared to most recent results of extraction of nPDFs using global QCD fits. Figure~\ref{fig:Comp_EPS_2017_v2}
shows a comparison of our predictions for $x g_A(x,Q_0^2)/[Axg_N(x,Q_0^2)]$ (the same as in Fig.~\ref{fig:Santi_sum_old_xdep_2014_comp}) to the EPPS16~\cite{Eskola:2016oht} (left panel) and nCTEQ15~\cite{Kovarik:2015cma} (right panel) results; the shaded error bands around the EPPS16 and nCTEQ15 curves quantify their uncertainties. 
 One can see from the left panel that while our predictions are within the EPPS16 uncertainty band, comparing to the 
 EPPS16 central curve one observes the trends of the 
 $x$ dependence are different. 
 In the right panel, the agreement between our predictions and the nCTEQ15 result is somewhat worse due to the flat and significant nuclear shadowing of $x g_A(x,Q_0^2)/[Axg_N(x,Q_0^2)]$ in the nCTEQ15 fit, which extends up to $x=0.01$ and thus leads to the large gluon antishadowing. 
Note that the shown uncertainties of the $x g_A(x,Q_0^2)/[Axg_N(x,Q_0^2)]$ ratio include only the nCTEQ15 nPDF errors.
  
It is important to emphasize that our approach is conceptually different from global QCD fits of nPDFs: while our 
predictions are based on microscopical dynamical models of nuclear shadowing and antishadowing, nPDFs extracted from
global QCD analyses present a model-dependent extrapolation for $x < 0.005$. In addition, theoretical uncertainties of
our results are significantly smaller than those of global QCD fits of nPDFs (see Fig.~\ref{fig:Comp_EPS_2017_v2})
and are largely controlled by a single parameter --- the effective $\sigma_{\rm soft}^j(x,Q_0^2)$ cross section in Eq.~(\ref{eq:shad_gen}).

\begin{figure}[h]
\begin{center}
\includegraphics[scale=1.4]{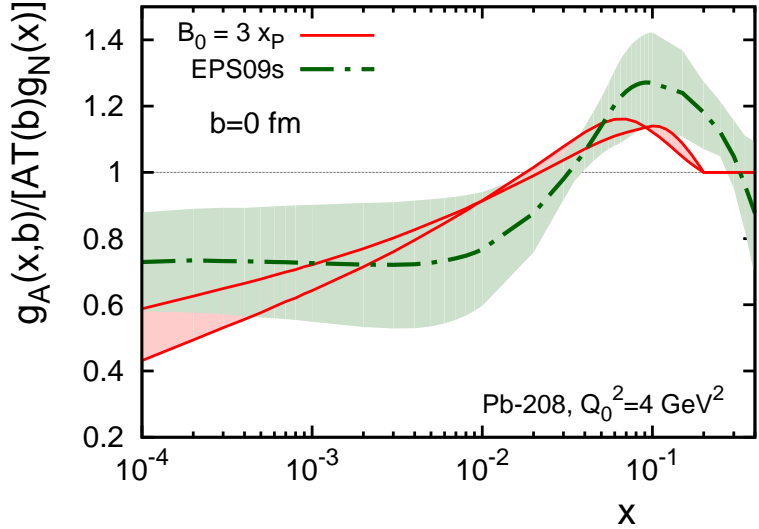}
\caption{Impact parameter dependent gluon nuclear shadowing and antishadowing.
Comparison of $x g_A(x,b,Q_0^2)/[AT_A(b)xg_N(x,Q_0^2)]$ predicted in the leading twist model of nuclear shadowing
and the dynamical model of antishadowing (same as in Fig.~\ref{fig:Santi_sum_old_xdep_2014_B}) to the EPS09s result.
The shaded areas show uncertainties of the respective predictions.}
\label{fig:Comp_EPS_bdep_2017}
\end{center}
\end{figure}

Figure~\ref{fig:Comp_EPS_bdep_2017} compares our predictions for the impact parameter dependence of nuclear shadowing 
and antishadowing in the $x g_A(x,b,Q_0^2)/[AT_A(b)xg_N(x,Q_0^2)]$ ratio to the EPS09s~\cite{Helenius:2012wd} result.
The shaded area around the EPS09s shows the fit uncertainty. 
One can see from the figure that the flat EPS09s nuclear shadowing extending up to $x=0.01$ requires the sizable 
gluon antishadowing. In our case, since antishadowing and shadowing compensate each other locally in rapidity, 
antishadowing noticeably reduces shadowing already for $x > 0.005$, which in turn does not require a very pronounced 
antishadowing enhancement. 

\section{Conclusions}
\label{sec:conclusions}

In this work, we explore the observation that in the infinite momentum frame, the nuclear effects of shadowing and antishadowing
originate from the same graph describing the merging of two parton ladders belonging to two different nucleons
of a nucleus and that this  merging is local in the rapidity. It enables us to 
propose that for a given momentum fraction $x_{\Pomeron}$ carried by the 
diffractive exchange, 
nuclear shadowing and antishadowing should compensate
each other in the momentum sum rule for nPDFs locally on the interval $\ln (x/x_{\Pomeron}) \le 1$.
This allows us to construct a model of nuclear gluon antishadowing, where it naturally has a wide support in 
$x$, $10^{-4} < x < 0.2$, peaks at $x=0.05-0.1$ and rather insignificantly depends on details of the model. 
In the studied example of the $x g_A(x,Q_0^2)/[Axg_N(x,Q_0^2)]$ ratio for $^{208}$Pb at $Q_0^2=4$ GeV$^2$,
our dynamical approach to antishadowing leads to $\approx 15$\% enhancement of this ratio at $x=0.05-0.1$. 
We also studied the impact parameter dependence of antishadowing and found it to be 
significantly slower 
that the $b$-dependence of the nuclear shadowing correction to nPDFs. 
While our predictions for the magnitude of nuclear shadowing and antishadowing of
the gluon nPDF agree in general to the EPPS16, EPS09s and nCTEQ15 results within their currently large uncertainties, 
the predicted shapes of the $x$ dependence are rather different. 
 
 \acknowledgments

The research of M.S. and L.F. was supported by the US Department of Energy Office of Science, Office of Nuclear Physics under Award 
No.~DE-FG02-93ER40771. V.G. would like to thank Pennsylvania State University for hospitality during the final stage of this work.

\end{document}